\documentclass[12pt]{article}
\usepackage{amsmath}
\usepackage{amsfonts}
\usepackage{amssymb}
\usepackage{bbold}
\usepackage{graphicx}%
\providecommand{\U}[1]{\protect\rule{.1in}{.1in}}
\begin{document}

\title{\textbf{Flat Directions and Leptogenesis in a ``New" $\mu \nu$SSM}}
\author{M. C. Rodriguez and I. V. Vancea\\\thanks{marcoscrodriguez@ufrrj.br, ionvancea@ufrrj.br} 
\emph{{\small Grupo de F{\'{\i}}sica Te\'{o}rica e Matem\'{a}tica
F\'{\i}sica, Departamento de F\'{\i}sica, }} \\\emph{{\small Universidade Federal Rural do Rio de Janeiro (UFRRJ),}} \\\emph{{\small Cx. Postal 23851, BR 465 Km 7, 23890-000 Serop\'{e}dica - RJ,
Brasil }} }
\date{05 July 2016}
\maketitle

\begin{abstract}
In this paper, we give a brief review of the Minimal Supersymmetric Standard Model (MSSM) and ``$\mu$ from $\nu$'' Supersymmetric Standard Model ($\mu \nu$SSM). Then we propose a generalization of $\mu \nu$SSM in order to explain the recent ATLAS, CMS and LHCb results. This ``new" $\mu \nu$SSM 
generalizes the superpotential $W_{suppot}$ of $\mu \nu$SSM by including two terms that generate a mixing among leptons, gauginos and higgsinos  while keeping the charginos and neutralinos masses unchanged. Also, it is potentially interesting for cosmological applications as it displays flat directions of the superpotential and a viable leptogenesis mechanism.
\end{abstract}

PACS number(s): 12.60. Jv


\vspace{2cm}

\newpage

\maketitle

\section{Introduction}

Despite its successful predictions, the Standard Model (SM) suffers from a major drawback as it contains massless neutrinos to all orders in the perturbation theory. Even after including the non-perturbative effects, this problem persists. This is in contradiction with the experimental results that suggest
that the neutrinos have non-zero masses and oscillations. The best-fit values at $1 \sigma$ error level for these neutrino oscillation parameters in the three-flavor framework are summarised as follows \cite{nu_best-fit}
\begin{eqnarray}
\Delta |m_{atm}^{2}|&=& \Delta |m^{2}_{31}| = 2.40^{+0.12}_{-0.11} \times 10^{-3} {\mbox eV}^{2} \;,\quad
\sin^{2} \theta_{atm}= \sin^{2} \theta_{12} = 0.304^{+0.022}_{-0.016}\;, \nonumber \\
\Delta m_{solar}^{2}&=& \Delta m^{2}_{21} = 7.65^{+0.23}_{-0.20} \times 10^{-5} {\mbox eV}^{2} \;,\quad
\sin^{2} \theta_{solar}=\sin^{2} \theta_{23} = 0.50^{+0.07}_{-0.06}\;, \nonumber \\
\sin^{2} \theta_{13} &\leq& 0.01^{+0.016}_{-0.011}\;. \nonumber \\
\label{eq1:best-fit_mass}
\end{eqnarray}
Thus, the oscillation experiments indicate that at least some neutrinos must be massive. However, the above relations do not provide the overall scale of masses which means that other methods must be employed to understand the neutrino mass spectrum.  One way to obtain meaningful bounds on the 
absolute scale for the neutrinos is to look for kinematic effects that can be present as a consequence of their non-zero masses in the tritium $\beta$-decay 
(${}^{3}$H $\rightarrow {}^{3}{\mbox He} + \overline{\nu}_{e} + e^{-}$). Two groups from Mainz \cite{mainz} and from Troitsk \cite{troitsk}, respectively, have reported on the bounds of $m_\nu < 2.3$~eV and $m_\nu < 2.5$~eV. Also, the upcoming KATRIN experiment \cite{katrin} is expected to produce results at a sensitivity of about 0.3~eV, which will further narrow down the scale of the neutrino spectrum. Another way to probe the neutrino mass scale is via studies of the lepton number ($L$) violating neutrinoless double $\beta$-decay 
($^{A}_{Z}\left[\mbox{Nucl}\right] 
\rightarrow \;^{\;\;\;\;A}_{Z+2}\left[\mbox{Nucl}^{\prime}\right] + 2 e^{-}$) \cite{bb_Maj_test}. Several groups such as Heidelberg-Moscow \cite{heidelberg-moscow} and IGEX \cite{IGEX} collaborations
conducted experiments with $^{76}{\mbox Ge}$, while the more recent CUORICINO experiment \cite{CUORICINO} used $^{130}{\mbox Te}$ to test the lepton number conservation. The best upper bounds on the decay lifetimes are presently provided by CUORICINO (which is still running), whose results are translated to
\begin{equation}
m_{\nu}< 0.19 -0.68 {\mbox eV } \;(90\% {\mbox C.L.}) \;,
\label{eq1:numass_CUOR}
\end{equation}
for the neutrino mass. Note that the large range is due to the uncertainty in the 
nuclear matrix elements. Upcoming experiments like CUORE \cite{CUORE}, GERDA \cite{GERDA} and MAJORANA \cite{MAJORANA} are expected to further improve these results with projected sensitivity of about 0.05~eV. Finally, it must be mentioned that some of the strongest bounds on the overall scale for neutrino masses come from 
cosmology. The studies of the data from the Wilkinson Microwave Anisotropy Probe (WMAP) and the Sloan Digital Sky 
Survey (SDSS) have deduced that the sum of neutrino masses (three species assumed) is constrained by $\sum_{i} \,
|m_{i}| \leq 0.6$ \cite{wmap_numass} and $1.6$~eV \cite{sdss_numass}. 

On the other hand, the LhCb reported recently a deviation of $2.6 \sigma$ of the measured ratios of the branching fractions $R_{K}$ in the individual lepton flavour model with respect to the SM in the low invariant mass region given by $1\ {\mbox GeV}^2 \leq M_{\ell\ell} \leq 6\ {\mbox GeV}^2$  \cite{Aaij:2014ora}. In this range, $R_{K}$ is defined as
\begin{equation}
R_K = 
\frac{\int^{q^{2}_{max}}_{q^{2}_{min}}\frac{d\Gamma(B^+ \rightarrow K^+ \mu^+ \mu^-)}{dq^2} dq^2}{\int^{q^{2}_{max}}_{q^{2}_{min}} \frac{d\Gamma(B^+ \rightarrow K^+ e^+ e^-)}{dq^2} dq^2}
\label{RK}
\end{equation}
Thus, the experimental results from \cite{Aaij:2014ora} put new numerical constraints on the scalar and pseudoscalar couplings. As observed in \cite{Aaij:2014ora,Aaij:2013pta}, the low invariant mass range excludes the resonant regions $J/\psi \rightarrow \mu^+ \mu^- $ and $J/\psi \rightarrow e^+ e^-$ thus improving the theoretical predictions. After these interesting results, the CMS collaboration published an intriguing deviation  from the SM in the $eejj$ channel, in the mass region $1.8$ {\mbox TeV}$< m_{eejj}<2.2 {\mbox TeV}$. No significant deviation was observed in the $\mu\mu jj$ channel \cite{Khachatryan:2014dka,CMS1}. The ATLAS measured an excess\footnote{The excess was at a 
di-boson invariant masses in the range from 1.3 to 3.0 TeV.} with respect to SM predictions in the production of 
di-electroweak gauge bosons VV (where V= W;Z) that decay hadronically \cite{Aad:2015owa}. The ATLAS and CMS 
collaborations have recently presented the results of di-photon resonance searches\footnote{The ressonance appear 
at around 750 GeV in the di-photon invariant mass.} in early Run II of $\sqrt s=13$ TeV data 
\cite{ATLAS-CONF-2015-081,CMS:2015dxe,atlas13,CMS:2016owr}.  

Beside the neutrino masses and the results from Atlas, LhCb and CMS, there are other features of the SM that require an explanation from a more fundamental point of view: 
\begin{enumerate}
\item The coupling constants do not meet at a single definite value \cite{amaldi}.
\item The hierarchy problem \cite{hier}.
\item The naturalness or fine tuning problem \cite{fine}.
\item The large number of parameters \cite{dress,Baer:2006rs,Aitchison:2005cf}.
\end{enumerate}
Also, it is expected that in a fundamental theory, the gravity take a natural place alongside the other three fundamental interactions. 

One promising class of theories that could solve the problems of the SM is formed by the supersymmetric extensions of the SM based on a postulated fundamental symmetry between the bosons and the fermions. The model from this class that contains a minimum number of physical states and interactions is the Minimal Supersymmetric Standard Model (MSSM) \cite{dress,Baer:2006rs,Aitchison:2005cf,R,ssm,grav}\footnote{About the history of MSSM, see e. g. \cite{Fayet:2001xk,Rodriguez:2009cd}}. The MSSM suffers from the $\mu$-problem which is the generation of a $\mu$ coupling in the $\mu \hat{H}_{1}\hat{H}_{2}$ term of the order of the electro-weak scale. The $\mu$-problem is solved by $\mu \nu$SSM proposed in \cite{LopezFogliani:2005yw}\footnote{The term $h^{i}_{\nu}\hat{H}_{2}\hat{H}_{1}\hat{\nu}^{c}_{i}$ generate the $\mu$ term when the sneutrino get its values expectation values.} which represents a modification of the MSSM by introducing new Yukawa interactions $Y_{\nu}^{ij} \, \hat{H}_{2}\, \hat{L}_{i} \, \hat{\nu}^{c}_{j}$ that generates light neutrino masses, as we will 
present at Sec.(\ref{sec:newsup}).  

The aim of the present paper is to propose a modification $\mu\nu$SSM that can explain the recent data from Atlas, CMS and LHCb by introducing new
interactions among the leptons with gauginos and higgsinos while the masses of charginos and neutralinos are left unchanged. Since the new model has an explicit broken $R$-parity and lepton number, there are flat directions of the superpotential that can generate the cosmological inflation. Also, the matter anti-matter asymmetry could be obtained from the letogenesis mechanism. 

This paper is organized as follows. In order to make the paper self-contained, we review in Section 2 the $\mu\nu$MSSM and establish our notations. In Section 3 we present a model that generalizes the $\mu\nu$MSSM . Next, we calculate all flat directions of this model and show that it can generate a viable leptogenesis mechanism
\footnote{We recall that the flat directions provide a viable mechanism to generate the cosmological inflation and the leptogenesis is important to explaining the asymmetry between the matter and the anti-matter.}. Also, we show how this model can explain the data from CMS and LHCb. The last section is devoted to conclusions.

\section{Review of the $\mu \nu$SMM }

The Minimal Supersymmetric Standard Model (MSSM) is the supersymmetric extension of the SM that contains a minimal number of states and interactions \cite{R,ssm,grav}. It aims at providing a general frame for the solving of the hierarchy problem, for the stabilization of the weak scale, for the unification of the coupling constants and for addressing the dark matter issues, among other things. The model has the
gauge symmetry $SU(3)_{C} \otimes SU(2)_{L} \otimes U(1)_{Y}$ extended by the supersymmetry to include the supersymmetric partners of the SM fields which have spins that differ by $+1/2$ as required by the supersymmetric algebra \cite{dress,Baer:2006rs,Aitchison:2005cf,Martin:1997ns}. Since the SM fermions are left-handed and right-handed and they transform differently under $SU(3)_{C}$, $SU(2)_{L}$ and $U(1)_{Y}$  groups, the particles of the MSSM must belong to chiral or gauge supermultiplets. The degrees of freedom are grouped in gauge superfields for gauge bosons and left-handed chiral superfields for spinors.

The chiral supermultiplet \cite{dress,Baer:2006rs,Aitchison:2005cf} contains three families of left-handed (right-handed) quarks
$\hat{Q}_{i}\sim({\bf 3},{\bf2},1/3)$ and $(\hat{u}^{c}_{i}\sim({\bf \bar{3}},{\bf1},-4/3)$, $\hat{d}^{c}_{i}\sim({\bf \bar{3}},{\bf1},2/3))$. Here, 
the numbers in parenthesis refers to the 
$(SU(3)_{C}, SU(2)_{L}, U(1)_{Y}$) quantum numbers, respectively and $i=1,2,3$ refers to the generation index (or flavor indices) and we neglected the color indices. Also, we use the notation in the anti-right-chiral superfield  
$e^{-}_{R}=(e^{+}_{L})^{c}$ according to \cite{Baer:2006rs}.
The model contains three families of leptons 
$\hat{L}_{i}\sim({\bf 1},{\bf2},-1)$ 
($\hat{l}^{c}_{i}\sim({\bf 1},{\bf1},2)$), respectively. The Higgs boson has spin 0, therefore it must belong to a chiral supermultiplet. However, in this case a single Higgs boson cannot provide mass for all quarks that have different weak isospin charges 
$T_{3} = \pm (1/2)$. Therefore, the MSSM contains two left-handed chiral superfields for Higgs fields $\hat{H}_{1}\sim({\bf 1},{\bf2},-1),\hat{H}_{2}\sim({\bf 1},{\bf \bar{2}},1)$ \cite{dress,Baer:2006rs,Aitchison:2005cf,Martin:1997ns,Kuroda:1999ks}. The particle content 
of each chiral superfield given above is presented in the Tables (\ref{lfermionnmssm}) and (\ref{rfermionnmssm}) below
\begin{table}[h]
\begin{center}
\begin{tabular}{|c|c|c|}
\hline 
$\mbox{ Left-Chiral Superfield} $ & $\mbox{ Fermion} $ & $\mbox{ Scalar} $ \\
\hline
$\hat{L}_{i}$ & $L_{i}$ & $\tilde{L}_{i}$ \\
\hline
$\hat{Q}_{i}$ & $Q_{i}$ & $\tilde{Q}_{i}$ \\ 
\hline
$\hat{H}_{1}$ & $\tilde{H}_{1}$ & $H_{1}$  \\
\hline
\end{tabular}
\end{center}
\caption{\small Particle content in the left-chiral superfields in MSSM, $i$ is flavour index ($i=1,2,3$).}
\label{lfermionnmssm}
\end{table}
\begin{table}[h]
\begin{center}
\begin{tabular}{|c|c|c|}
\hline 
$\mbox{ Anti-Right-Chiral Superfield} $ & $\mbox{ Fermion} $ & $\mbox{ Scalar} $ \\
\hline
$\hat{l}^{c}_{i}$ & $l^{c}_{i}$ & $\tilde{l}^{c}_{i}$ \\ 
\hline
$\hat{u}^{c}_{i}$ & $u^{c}_{i}$ & $\tilde{u}^{c}_{i}$ \\ 
\hline
$\hat{d}^{c}_{i}$ & $d^{c}_{i}$ & $\tilde{d}^{c}_{i}$ \\ 
\hline
$\hat{H}_{2}$ & $\tilde{H}_{2}$ & $H_{2}$   \\
\hline
\end{tabular}
\end{center}
\caption{\small Particle content in the anti-right-chiral superfields in MSSM, $i$ is flavour index ($i=1,2,3$).}
\label{rfermionnmssm}
\end{table}

The gauge supermultiplets are described by three vector superfilds $\hat{V}^{a}_{c}\sim({\bf 8},{\bf 1}, 0)$, where $a=1,2, \ldots ,8$, $\hat{V}^{i}\sim({\bf 1},{\bf 3}, 0)$ with $i=1,2,3$ and 
$\hat{V}^{\prime}\sim({\bf 1},{\bf 1}, 0)$. The particle content 
in each vector superfield is presented in the Table \ref{gaugemssm}.
\begin{table}[h]
\begin{center}
\begin{tabular}{|c|c|c|c|}
\hline 
${\rm{Vector \,\ Superfield}}$ & ${\rm{Gauge \,\ Bosons}}$ & ${\rm{Gaugino}}$ & ${\rm Gauge \,\ constant}$ \\
\hline 
$\hat{V}^{a}_{c}$ & $g^{a}$ & $\tilde{g}^{a}$ & $g_{s}$ \\
\hline 
$\hat{V}^{i}$ & $V^{i}$ & $\tilde{V}^{i}$ & $g$ \\
\hline
$\hat{V}^{\prime}$ & $V^{\prime}$ & $\tilde{V}^{\prime}$ & $g^{\prime}$ \\
\hline
\end{tabular}
\end{center}
\caption{\small Particle content in the vector superfields in MSSM.}
\label{gaugemssm}
\end{table}

The supersymetric Lagrangian of the MSSM is given by
\begin{equation}
\mathcal{L}_{SUSY} = \mathcal{L}^{chiral}_{SUSY} + \mathcal{L}^{Gauge}_{SUSY} .
\label{SUSY-Lagrangian1}
\end{equation}
The Lagrangian defined in the equation (\ref{SUSY-Lagrangian1}) contains contributions from all sectors of the model
\begin{equation}
\mathcal{L}^{chiral}_{SUSY} = 
\mathcal{L}_{Quarks} + \mathcal{L}_{leptons} + \mathcal{L}_{Higgs} ,
\label{SUSY-Lagrangian}
\end{equation}
and the terms have the following explicit form
\begin{eqnarray}
{\cal L}_{Quarks}&=& \int d^{4}\theta\;\sum_{i=1}^{3}\left[\,
\hat{\bar{Q}}_{i}e^{2g_{s}\hat{V}_{c}+2g\hat{V}+g^{\prime} \left( \frac{1}{6} \right) \hat{V}^{\prime}} \hat{Q}_{i} +
{\hat{ \bar{u^c}}}_{i}e^{2g_{s}\hat{V}_{c}+ g^{\prime} \left( - \frac{2}{2}\right) \hat{V}^{\prime}}\hat{u}^{c}_{i} +
\hat{ \bar{d^{c}}}_{i}e^{2g_{s}\hat{V}_{c}+g^{\prime}\left( \frac{1}{3}\right) \hat{V}^{\prime}}\hat{d}^{c}_{i}
\,\right] \,\ . 
\end{eqnarray}
here, $\hat{V}_{c}=T^{a}\hat{V}^{a}_{c}$ and $T^{a}=\lambda^{a}/2$ (with $a=1,\cdots,8$) are the generators of $SU(3)_{C}$ and $\hat{V}=T^{i}\hat{V}^{i}$ where $T^{i}=\lambda^{i}/2$ (with $i=1,2,3$) are the
generators of $SU(2)_{L}$. As usual, $g_{s}$, $g$ and $g^{\prime}$ are the gauge couplings for the $SU(3)$, $SU(2)$ and $U(1)$ groups, respectively, as shown in the Table \ref{gaugemssm}. The action in the lepton and Higgs sectors are defined by the following Lagrangians 
\begin{eqnarray}
{\cal L}_{lepton}&=& \int d^{4}\theta\;\sum_{i=1}^{3}\left[\,
\hat{ \bar{L}}_{i}e^{2g\hat{V}+g^{\prime} \left( - \frac{1}{2}\right) \hat{V}^{\prime}} \hat{L}_{i} +
\hat{ \bar{l^{c}}}_{i}e^{g^{\prime}\hat{V}^{\prime}} \hat{l}^{c}_{i}\,\right] \,\ , \nonumber \\
{\cal L}_{Higgs}&=&  \int d^{4}\theta\;\left[\,
\hat{ \bar{H}}_{1}e^{2g\hat{V}+g^{\prime}\left( - \frac{1}{2}\right) \hat{V}^{\prime}}\hat{H}_{1}+
\hat{ \bar{H}}_{2}e^{2g\hat{V}+g^{\prime}\left( \frac{1}{2}\right) \hat{V}^{\prime}}\hat{H}_{2}
+  W+ \bar{W} \right]\!. 
\label{allsusyterms}     
\end{eqnarray}
The last two terms define the superpotential of the MSSM as $W=W_{H}+W_{Y}$ where
\begin{eqnarray}
W_{H}&=& \mu\; \epsilon_{\alpha \beta}\hat{H}_{1}^{\alpha}\hat{H}_{2}^{\beta} \,\ , 
\label{mu-HH} \\
W_{Y}&=& \epsilon_{\alpha \beta}\sum_{i,j=1}^{3}\left[\,
f^{l}_{ij}\hat{H}^{\alpha}_{1}\hat{L}^{\beta}_{i}\hat{l}^{c}_{j}+
f^{d}_{ij}\hat{H}^{\alpha}_{1}\hat{Q}^{\beta}_{i}\hat{d}^{c}_{j}+
f^{u}_{ij}\hat{H}^{\alpha}_{2}\hat{Q}^{\beta}_{i}\hat{u}^{c}_{j}\,\right] \,\ . 
\label{Y-H}               
\end{eqnarray}
The supersymmetric  parameter $\mu$ is a complex numbers and the $f$ terms are elements of the complex  
$3 \times 3$ Yukawa coupling matrices in the family space. The color indices on the triplet (antitriplet) 
superfield $\hat{Q}$ $( \hat{u}^{c}, \hat{d}^{c})$ contract trivially, and have been suppressed. The second terms of the Lagrangian defined by the equation (\ref{SUSY-Lagrangian1}) is given by the following equation
\begin{eqnarray}
\mathcal{L}^{Gauge}_{SUSY}&=&  \frac{1}{4} \int  d^{2}\theta\;
\left[ \sum_{a=1}^{8} W^{a \alpha}_{s}W_{s \alpha}^{a}+ \sum_{i=1}^{3} W^{i \alpha}W_{ \alpha}^{i}+
W^{ \prime \alpha}W_{ \alpha}^{ \prime}\, + h.c. \right] \,\ . \nonumber 
\end{eqnarray}
The gauge superfields have the following explicit form
\begin{eqnarray}
W^{a}_{s \alpha}&=&-\frac{1}{8g_{s}}\,\bar{D}\bar{D}e^{-2g_{s}\hat{V}^{a}_{c}}D_{\alpha}e^{2g_{s}\hat{V}^{a}_{c}} 
\,\ , \nonumber \\
W^{i}_{\alpha}&=&-\frac{1}{8g}\,\bar{D}\bar{D}e^{-2g\hat{V}^{i}}D_{\alpha}e^{2g\hat{V}^{i}} 
\,\ , \nonumber \\
W_{\alpha}^{\prime}&=&-\frac{1}{4}\,D D \bar{D}_{\alpha} \hat{V}^{\prime} \,\ ,
\label{W-a}
\end{eqnarray}
where $\alpha = 1,2$ is a spinorial index.

In principle, one could add to the Lagrangian defined by the equation (\ref{SUSY-Lagrangian}) other terms that, even if they break the baryon number and the lepton number conservation laws, are still allowed by the supersymmetry. However, no physical process with this property has been discovered so far. This phenomenological fact suggest imposing a symmetry that rules out such terms called $R$-parity which is defined in terms of the following operators
\begin{equation}
P_{M} = (-1)^{3(B-L)}, \hspace{0,5cm}
P_{R} = P_{M}(-1)^{2s},
\end{equation}
where $B$ and $L$ are the baryon and lepton numbers, respectively, and $s$ is the spin for a given state. The $R$-parity of the Lagrangian implies that the usual particles of the SM have $P_{M}=1$ while their superpartners have $P_{M}=-1$. 
The terms that break $R$-parity are  
\begin{eqnarray}
W_{2RV}&=&\epsilon_{\alpha \beta} \sum_{i=1}^{3}\mu_{0i} \hat{L}^{\alpha}_{i}\hat{H}^{\beta}_{2},\nonumber \\
W_{3RV}&=&\epsilon_{\alpha \beta} \sum_{i,j,k=1}^{3} \left(  \lambda_{ijk}\hat{L}^{\alpha}_{i}\hat{L}^{\beta}_{j}\hat{l}^{c}_{k}+
\lambda^{\prime}_{ijk}\hat{L}^{\alpha}_{i}\hat{Q}^{\beta}_{j}\hat{d}^{c}_{k}+ 
\lambda^{\prime\prime}_{ijk}\hat{u}^{c}_{i}\hat{d}^{c}_{j}\hat{d}^{c}_{k} \right) .
\label{mssmrpv}
\end{eqnarray}
Here, we have suppressed the $SU(2)$ indices and $\epsilon$ is the
antisymmetric $SU(2)$ tensor. Some of the coupling constants in the equation (\ref{mssmrpv}) should be
set to zero in order to avoid a too fast proton decay and neutron-anti-neutron oscillation 
\cite{dress,Baer:2006rs,barbier,moreau}. The choice of the $R$-parity violation couplings $\lambda^{\prime}_{11k}$ with $k = 2$ and 3, are constrained from various low energy observables such as: (i) charge-current universality, (ii) $e-\mu-\tau$ universality, (iii) atomic parity violation. 
The bounds on the product $|\lambda^{\prime}_{112}\lambda^{\prime}_{113}|$ can be obtained from the charged $B$-meson decay mixing $B^{\pm}_d \rightarrow \pi^{\pm} K^0$, $B_{s}-\bar{B}_{s}$ and the transition  $B\rightarrow X_s\gamma$ as observed in \cite{Biswas:2014gga}. 

The experimental evidence suggests that the supersymmetry is not an exact symmetry. Therefore, supersymmetry 
breaking terms should be added to the Lagrangian defined by the equation (\ref{SUSY-Lagrangian}). One possibility is by 
requiring that the divergences cancel at all orders of the perturbation theory. The most general soft supersymmetry 
breaking terms, which do not induce quadratic divergence, where described by Girardello and Grisaru \cite{10}. They 
found that the allowed terms can be categorized as follows: a scalar field $A$ with mass terms
\begin{equation}
{\cal L}_{SMT}=-m^{2} A^{\dagger}A,
\end{equation}
 a fermion field gaugino  $\lambda$ with mass  terms
\begin{equation}
{\cal L}_{GMT}=- \frac{1}{2} (M_{ \lambda} \lambda^{a} \lambda^{a}+h.c.)
\end{equation}
and finally trilinear scalar interaction terms
\begin{equation}
{\cal L}_{INT}= \Xi_{ij}A_{i}A_{j}+ \Upsilon_{ij}A_{i}A_{j}+ \Omega_{ijk}A_{i}A_{j}A_{k}+h.c.
\end{equation}
The terms in this case are similar with the terms allowed in the superpotential of the model we are going to consider next.

Taken all this information into account, we can add the following soft supersymmetry breaking terms to the MSSM
\begin{eqnarray}
{\cal L}^{MSSM}_{Soft} &=& {\cal L}^{MSSM}_{SMT} + {\cal L}^{MSSM}_{GMT}+ {\cal L}^{MSSM}_{INT} \,\ ,
\label{The Soft SUSY-Breaking Term prop 2aaa}
\end{eqnarray}
where the scalar mass term ${\cal L}_{SMT}$ is given by the following relation
\begin{eqnarray}
{\cal L}^{MSSM}_{SMT} &=& - \sum_{i,j=1}^{3} \left[\,
\left( M_{L}^{2}\right)_{ij}\;\tilde{L}^{\dagger}_{i}\tilde{L}_{j}+ 
\left( M^{2}_{l}\right)_{ij} \tilde{l^{c}}^{\dagger}_{i}\tilde{l^{c}}_{j}+ 
\left( M_{Q}^{2}\right)_{ij}\;\tilde{Q}^{\dagger}_{i}\tilde{Q}_{j}
\right. \nonumber \\  
\hspace{1.7cm} &+& \left.
\left( M^{2}_{u}\right)_{ij} \tilde{u^{c}}^{\dagger}_{i}\tilde{u^{c}}_{j}+ 
\left( M^{2}_{d}\right)_{ij} \tilde{d^{c}}^{\dagger}_{i}\tilde{d^{c}}_{j}
+ M_{1}^{2} H^{\dagger}_{1}H_{1} +
M_{2}^{2} H^{\dagger}_{2}H_{2}
 \right] \,\ ,
      \label{burro}
\end{eqnarray}
The $3 \times 3$ matrices $M_{L}^{2},M^{2}_{l},M_{Q}^{2},M^{2}_{u}$ and $M^{2}_{d}$ are hermitian and $M_{1}^{2}$ and $M_{2}^{2}$ are real. The gaugino mass term is written as
\begin{eqnarray}
{\cal L}^{MSSM}_{GMT} &=&- \frac{1}{2}  \left[
\left(\,M_{3}\; \sum_{a=1}^{8} \lambda^{a}_{C} \lambda^{a}_{C}
+ M\; \sum_{i=1}^{3}\; \lambda^{i}_{A} \lambda^{i}_{A}
+ M^{\prime} \;   \lambda_{B} \lambda_{B}\,\right)
+ h.c. \right] \,\ .
\label{The Soft SUSY-Breaking Term prop 3}
\end{eqnarray}
Here, $M_{3},M$ and $M^{\prime}$ are complex. Finally, there is an interaction term ${\cal L}_{INT}$, see the equation (\ref{mssmrpv}), of the form
\begin{eqnarray}
{\cal L}^{MSSM}_{INT} =- M_{12}^{2}\epsilon H_{1}H_{2} 
+ \epsilon \sum_{i,j,k=1}^{3} \left[ 
\left( A^{E}\right)_{ij} H_{1}\tilde{L}_{i}\tilde{l}^{c}_{j}+
\left( A^{D}\right)_{ij} H_{1}\tilde{Q}_{i}\tilde{d}^{c}_{j}+
\left( A^{U}\right)_{ij} H_{2}\tilde{Q}_{i}\tilde{u}^{c}_{j}\right] +h.c.  \,\ .
\end{eqnarray} 
The $3 \times 3$ matrices $M_{12}^{2}$ and $A$ matrices are complex. 

The total Lagrangian of the MSSM is obtained by adding all Lagrangians above
\begin{equation}
\mathcal{L}^{MSSM} = \mathcal{L}_{SUSY} + \mathcal{L}^{MSSM}_{soft},
\label{L-total}
\end{equation}
see the equations (\ref{SUSY-Lagrangian1},\ref{The Soft SUSY-Breaking Term prop 2aaa}). The MSSM contains 124 free 
parameters \cite{Baer:2006rs} and the symmetry breaking parameters are completely arbitrary \cite{dress}. The main goal in the SUSY phenomenology is to find some approximation about the way we can break SUSY in order 
to have a drastic reduction in the number of these parameters\footnote{Different assumptions result in 
different version of the Constrained Minimal Supersymmetric Model (CMSSM).}. Many phenomenological analyses adopt 
the universality hypothesis at the scale $Q \simeq M_{GUT}\simeq 2 \times 10^{16}$ GeV:
\begin{eqnarray}
g_{s}&=&g=g^{\prime} \equiv g_{GUT}, \nonumber \\
M_{3}&=&M=M^{\prime}\equiv m_{1/2}, \nonumber \\
M_{L}^{2}&=&M^{2}_{l}=M_{Q}^{2}=M^{2}_{u}=M^{2}_{d}=M_{1}^{2}=M_{2}^{2}\equiv m^{2}_{0}, \nonumber \\
A^{E}&=&A^{D}=A^{U}\equiv A_{0}.
\label{msugra}
\end{eqnarray}
The assumptions that the MSSM is valid between the weak scale and GUT scale, and that the "boundary conditions", 
defined by  the equation (\ref{msugra}) hold, are often referred to as mSUGRA, or minimal supergravity model. The mSUGRA 
model is completely specified by the parameter set \cite{dress,Baer:2006rs}
\begin{eqnarray}
m_{0}, \,\ m_{1/2}, \,\ A_{0}, \,\ \tan \beta , \,\ {\mbox sign}( \mu ) .
\end{eqnarray}
The new free parameter $\beta$ is defined in the following way
\begin{eqnarray}
\tan \beta \equiv \frac{v_{2}}{v_{1}},
\label{defbetapar}
\end{eqnarray}
where $v_{2}$ is the vev of $H_{2}$ while $v_{1}$ is the vev of the Higgs in the doublet representation 
of $SU(2)$ group. Due the fact that $v_{1}$ and $v_{2}$ are both positive, it imples that 
$0 \leq \beta \leq (\pi/2) \,\ {\mbox rad}$.

In the context of the MSSM, it is possible to give mass to all charged fermions. With this superpotential we can explain the mass hierarchy in the charged fermion masses as showed in \cite{cmmc,cmmc1}. On the other hand, ${\cal L}_{Higgs}$ give mass to the gauge bosons: the charged ones ($W^{\pm}$) and the neutral ($Z^{0}$ and get a massless foton but the neutrinos remain massless. Due to this fact, it is generated a spectrum that contains five physical Higgs bosons, two neutral scalar ($H,h$), one neutral pseudoscalar ($A$), and a pair of charged Higgs particles ($H^{\pm}$). At the level of tree 
level, we can write the following relations hold in the Higgs sector \cite{dress,Baer:2006rs}:
\begin{eqnarray}
m^{2}_{H^{\pm}}&=&m^{2}_{A}+m^{2}_{W}, \nonumber \\
m^{2}_{h,H}&=&\frac{1}{2}\left[ 
(m^{2}_{A}+m^{2}_{Z}) \mp \sqrt{(m^{2}_{A}+m^{2}_{Z})^{2}-4m^{2}_{A}m^{2}_{Z}\cos^{2}\beta} \right], \nonumber \\
m^{2}_{h}+m^{2}_{H}&=&m^{2}_{A}+m^{2}_{Z}, \nonumber \\
\cos^{2}( \beta - \alpha )&=& \frac{m^{2}_{h}(m^{2}_{Z}-m^{2}_{h})}{m^{2}_{A}(m^{2}_{H}-m^{2}_{h})}. 
\end{eqnarray}
Therefore, the light scalar $h$ has a mass smaller than the $Z^{0}$ gauge boson at the tree level. This implies that one has to 
consider the one-loop corrections which lead to the following result \cite{Haber:1990aw}
\begin{equation}
m_{h}\simeq m^{2}_{Z}+ \frac{3g^{2}m^{4}_{Z}}{16 \pi^{2}m^{2}_{W}}\left\{ 
\ln \left( \frac{m^{2}_{\tilde{t}}}{m^{2}_{t}}\right) \left[ \frac{2m^{4}_{t}-m^{2}_{t}m^{2}_{Z}}{m^{4}_{Z}}\right] +
\frac{m^{2}_{t}}{3m^{2}_{Z}}\right\}.
\end{equation}
In the MSSM there are four neutralinos ($\tilde{\chi}^{0}_{i}$ with $i=1,2,3,4$) and two charginos 
($\tilde{\chi}^{\pm}_{i}$ with $i=1,2$) \cite{dress,Baer:2006rs}.

The mass matrix of neutrinos from this model was studied in \cite{hall,banks,rv1,fb}. The mass matrix has two zero eigenvalues. Thus, there are two neutrinos $\nu_{1,2}$, which are massless at the tree level. More realistic neutrino masses require radiative corrections \cite{rv1,rnm,rv2,marta}. The neutrinos are Majorana particles, therefore the neutrinoless double beta decay must be observed.
Neutrinoless double beta decay $0\nu\beta\beta$ is a sensitive probe of physics beyond the SM since it violates the conservation of the lepton number \cite{hdmo94,baudis97,heidel01,arxivnote,expbb}. The nucleon level process of $0\nu\beta\beta$ decay, $ n +n \to p+p + e^{-} + e^{-}$, can be obtained via the lepton number violating sub-process $d+d \to u +u +e+e$. 

The supersymmetric mechanism of
$0\nu\beta\beta$ decay was first suggested by Mohapatra
\cite{rnm} and further studied in 
\cite{Vergados,HKK1}. In \cite{HKK2}, the $R$-parity
violating Yukawa coupling of the first generation is strongly
bounded by $\lambda^{\prime}_{111} \leq 3.9\cdot 10^{-4}$ due to the
gluino exchange $0\nu\beta\beta$-decay. Babu and Mohapatra
\cite{BM} have latter implemented another contribution comparable
with that via the gluino exchange. This set stringent bounds on
the products of $R$-parity violating Yukawa couplings
$\lambda^{\prime}_{11i}\lambda^{\prime}_{1i1}$ of $i$th generation index \cite{hir96} 
\begin{eqnarray}
\lambda_{113}^{\prime}\lambda_{131}^{\prime} &\leq & 1.1 \cdot 10^{-7},\\
\lambda_{112}^{\prime}\lambda_{121}^{\prime}&\leq & 3.2 \cdot 10^{-6}. 
\end{eqnarray}

On the other hand, the confrontation of the experimental results with the predictions of the MSSM set a phenomenological constraint on the magnitude of 
$\mu \sim {\cal O}(M_{W})$. Indeed, the mass of Higgssino from the equation (\ref{mu-HH}) and the terms from the 
${\cal L}_{Soft}$, given by the equation (\ref{The Soft SUSY-Breaking Term prop 2aaa}), are of order of the electro-weak scale of $246 \,\ GeV$ while the natural cut-off scale is the Planck scale $1.22 \times 10^{19}\,\ GeV$. The MSSM does not provide any mechanism to explain the difference between the two scales. This is know as the \emph{$\mu$ problem}. 

In order to address this problem, the Next-to-the-Minimal Supersymmetric Standard-Model (NMSSM) \cite{R,nmssm} was developed within the framework of the Grand Unification Theory (GUTs) as well as the superstring theorie \cite{barr,nilsredwy,derendinger}\footnote{References to the original work on the NMSSM may be found in the reviews \cite{Ananthanarayan:1996zv,Ellwanger:1998jk}}. The NMSSM is characterized by the a new singlet field introduced in the following chiral superfield\footnote{$y^{m}\equiv x^{m}-i \theta \sigma^{m}\bar{\theta}$, 
where $\sigma^{m}$ are the three Pauli matrices plus the $I_{2 \times 2}$ the identity matrix.} \cite{dress}
\begin{eqnarray}
\hat{N}(y, \theta )&=&n(y)+ \sqrt{2}\theta \tilde{n}(y)+ \theta \theta F_{n}(y), 
\end{eqnarray}
where $n$ is the scalar in the singlet and its vacuum expectation value is given by $\sqrt{2} \langle n \rangle =x$. Its superpartner  $\tilde{n}$ is known as the singlino. The rest of the particle content of this model is the same as of the MSSM given above in the Tables (\ref{lfermionnmssm}), (\ref{rfermionnmssm} and \ref{gaugemssm}). The 
superpotential of the NMSSM model has the following form
\begin{eqnarray}
W_{NMSSM}&=&W_{Y}+ \epsilon_{\alpha \beta}\lambda \hat{H}_{1}^{\alpha} \hat{H}_{2}^{\beta} \hat{N}+
\frac{1}{3} \kappa \hat{N}\hat{N}\hat{N},
\end{eqnarray}
where $W_{3RV}$ is defined by the equation (\ref{Y-H}). The way in which the $\mu$-problem is solved in the NMSSM is 
by generating 
dynamically the $\mu$ term in the superpotential through $\mu = \lambda x$ with a dimensionless coupling $\lambda$ 
and the vacuum expectation value $x$ of the Higgs singlet. Another essential feature of the NMSSM is the fact that 
the mass bounds for the Higgs bosons and neutralinos are weakened. For more details about the scalar sector of this 
model see \cite{Drees:1988fc}. We summarize them in the Table \ref{tabvec} below 
\begin{table}[h]
\begin{center}
\begin{tabular}{|c|c|}
\hline
${\rm{Symbol}}$ &  ${\rm{Decomposition}}$  \\
\hline
$H^{\pm}$ &  $\sin(\beta) h_{1}^{\pm}+ \cos(\beta) h_{2}^{\pm}$  \\
\hline
$A_{1}$, $A_{2}$, $m_{A_{1}}\leq m_{A_{2}}$ 
& 
$A_{1}= \cos( \alpha_{PS})a^{0}+ \sqrt{2}\sin( \alpha_{PS}) \mathfrak{Im} \left[ n \right]$ 
\\
& $A_{2}=- \sin( \alpha_{PS})a^{0}+ \sqrt{2}\cos( \alpha_{PS}) \mathfrak{Im} \left[ n \right]$  \\
\hline
$h_{1}$, $h_{2}$, $h_{3}$, $m_{h_{1}}\leq m_{h_{2}}\leq m_{h_{3}}$ 
& 
$h_{i}= \sqrt{2} \mathfrak{Re}  \left[ {\cal O}_{i1}(h^{0}_{1}-v_{1})+ {\cal O}_{i2}(h^{0}_{2}-v_{2}) 
+ {\cal{O}}_{i3}(n-x) \right]$ \\
\hline
\end{tabular}
\end{center}
\caption{\small The physical Higgs states of the NMSSM \cite{dress}, the $\beta$ parameter is defined by the equation (\ref{defbetapar}).}
\label{tabvec}
\end{table}
Note that the neutralino sector is extended to a $5 \times 5$ mass matrix. If the following vector basis for fields is 
adopted (see, e. g. \cite{dress})
\begin{equation}
(\psi^{0})^{T}=(\lambda_{0},\lambda_{3},\tilde{h}^{1}_{1},\tilde{h}^{2}_{2},\tilde{n}) ,
\end{equation}
the mass matrix takes the following form
\footnote{Where we have defined $e=g \sin \theta_{W}=g^{\prime}\cos \theta_{W}$.}
\begin{equation}
Y= \left(
\begin{array}{ccccc}
M_{1} & 0 & -m_{Z}\sin \theta_{W} \cos \beta & m_{Z}\sin \theta_{W}\sin \beta & 0 \\
0 & M_{2} & m_{Z}\cos \theta_{W}\cos \beta & -m_{Z}\cos \theta_{W}\sin \beta & 0 \\
-m_{Z}\sin \theta_{W}\cos \beta & m_{Z}\cos \theta_{W}\cos \beta & 0 & - \lambda \frac{x}{\sqrt{2}} & 
- \lambda \frac{v_{1}}{\sqrt{2}} \\
m_{Z}\sin \theta_{W}\sin \beta & m_{Z}\cos \theta_{W}\sin \beta & - \lambda \frac{x}{\sqrt{2}} & 0 & 
- \lambda \frac{v_{2}}{\sqrt{2}} \\
0 & 0 & - \lambda \frac{v_{1}}{\sqrt{2}} & - \lambda \frac{v_{2}}{\sqrt{2}} & \sqrt{2}\kappa x
\end{array}
\right) .
\end{equation}
We note that the singlino $\tilde{n}$ does not mix directly with the gauginos $\lambda_{0},\lambda_{3}$ but it 
can mix with the neutral higgsinos \cite{dress}. The neutrinos are massless. 
However, the term $\hat{\nu}^{c}_{i} \hat{H}_{1}\hat{H}_{2}$ can produce an effective  $\mu$ term when the sneutrinos get vev as we will show later on. This would allow us to solve the $\mu$ problem \cite{mupb}, without having to introduce an extra singlet superfield as we have done in the NMSSM. This new model is called "$\mu$ from $\nu$''
Supersymmetric Standard Model ($\mu \nu$SSM). The field content of the $\mu \nu$SSM is the same as MSSM supplemented by three neutrino superfields $\hat{\nu}^{c}_{i}$ \cite{LopezFogliani:2005yw} and is given by the following equation
given by:
\begin{eqnarray}
\hat{\nu}^{c}_{i}(y, \theta )= \tilde{\nu}^{c}_{i}(y)+ \sqrt{2}\theta \nu^{c}_{i}(y)+ \theta \theta 
F_{\nu^{c}_{i}}(y).
\end{eqnarray} 
If the terms like $\hat{H}_{2}\hat{L}_{i}\hat{\nu}^{c}_{j}$ are considered, then the term $\mu_{0i}$ is induced 
when the right handed sneutrinos acquires a vev. By adding right handed neutrinos to the model, one can choose only the terms that break the lepton number conservation instead of the ones that break the baryon number conservation. The vev of these models are 
\begin{eqnarray}
\langle \tilde{\nu}^{c}_{i} \rangle &\equiv& \frac{v_{\nu^{c}_{i}}}{\sqrt{2}}, \nonumber \\
\langle \tilde{\nu}_{i} \rangle &\equiv& \frac{v_{\nu_{i}}}{\sqrt{2}}.
\label{vevemunussm}
\end{eqnarray} 
In this case, all the neutrinos of the model can 
get mass at the tree level. Then the double beta decay can occur and the nucleon is stabilized.

The ${\cal Z}_{3}$ symmetry generates the following transformation of each chiral superfield
\begin{equation}
\Phi \rightarrow \exp \left( \frac{2 \pi \omega}{3} \right) \Phi ,
\end{equation}
where $\omega$ is an entire number. The superpotential of this model can be obtained by requiring that it be ${\cal Z}_{3}$-symmetric invariant. As a consequence, it takes the following form   
\begin{eqnarray}
W_{\mu \nu suppot} &=&W_{Y}+ \sum_{i,j,k=1}^{3} \left( 
f^{\nu}_{ij} \, \hat{H}_{2}\, \hat{L}_{i} \, \hat{\nu}^{c}_{j} +
h_{i}^{\nu}\hat{H}_{2}\hat{H}_{1}\hat{\nu}^{c}_{i}
+ \frac{1}{3}\kappa^{ijk} \hat{\nu}^{c}_{i}\hat{\nu}^{c}_{j}\hat{\nu}^{c}_{k} \right) \,.
\label{superpotential}
\end{eqnarray}
where $W_{Y}$ is defined by the equation (\ref{Y-H}). It turns out that ${\cal Z}_{3}$ symmetry forbids all the bilinear terms 
in the superpotential. The expression obtained in the equation (\ref{superpotential}) is consistent with the phenomenological models derived from the superstring theory that generate only trilinear couplings. 

In the present context the string theory is relevant to the unification of all interactions, including gravity. The term proportional to $\kappa$ gives an effective Majorana mass term to neutrinos, while the coupling 
$f^{\nu}$ generates Dirac mass term to neutrinos. 

When the scalar components of the superfields $\hat{\nu}^{c}_{i}$, denoted by $\tilde{\nu}^{c}_{i}$, acquire
vev's of the order of the electroweak scale, an effective interaction $\mu \hat{H}_{1} \hat{H}_{2}$ is generated
with the effective coupling $\mu$ given by  
\begin{equation}
\mu \equiv h_{i}^{\nu} \langle \tilde{\nu}^{c}_{i} \rangle .
\end{equation} 
In the same situation, the term $\mu_{0i}\hat{H}_{2}\hat{L}_{i}$ can be generated with 
\begin{equation}
\mu_{0i}\equiv \sum_{j=1}^{3} f^{\nu}_{ij}\langle \tilde{\nu}^{c}_{j} \rangle ,
\end{equation}
the contribution of $f^{\nu}\leq 10^{-6}$ to the minimization conditions for the left-handed neutrinos 
$\mu_{0i}\ll \mu$. That provides an explanation for the neutrino's masses in MSSM \cite{Montero:2001ch}. 

In this model the $R$-parity (and also the lepton number conservation) is broken explicitly. One of the candidates for the dark matter in NMSSM is the gravitino. Recently, some experimental bounds on 
gravitino masses were presented in see \cite{Catena:2014pca}. For an analysis of the gravitino as dark matter without $R$-parity see \cite{yamaguchi}. Other possibilities that LSP be the axino were presented in \cite{axino}. The mass spectrum of this model can be found in \cite{Escudero:2008jg} and ths spectrum is consistent with the experimental values obtained for both masses and mixing. The nice phenomenological aspects of this model were discussed in \cite{Munoz:2009an}. There are some works in $\mu\nu$SSM that consider gravitino as dark matter \cite{Choi:2009ng,GomezVargas:2011ph,Albert:2014hwa}.

\section{A ``new" $\mu\nu$SSM}

In this section we propose a generalization of the $\mu\nu$SSM by adding new interaction terms that explicitly break the $R$-parity and the lepton number symmetries, respectively. Therefore, the new model has potentially interesting cosmological consequences such as flat directions that provide a mechanism for the cosmological inflation and leptogenesis which explains the asymmetry between the matter and the anti-matter. We determine the flat directions of the generalized $\mu\nu$SSM and explain the leptogenesis mechanism. Then we show how our proposal addresses the recent experimental results from CMS and LHCb obtained in \cite{Aaij:2014ora,Aaij:2013pta} and discussed in \cite{Biswas:2014gga}. 

\subsection{New terms in the Superpotential of $\mu \nu$SSM Model}
\label{sec:newsup}

The superpotential of the $\mu\nu$SSM model given by the equation (\ref{superpotential}) can be generalized as follows
\begin{eqnarray}
W &=& W_{\mu \nu suppot}+ \sum_{i,j,k=1}^{3} \left( \lambda^{\prime}_{ijk} \hat{L}_{i}\hat{L}_{j}\hat{l}^{c}_{k}+ \lambda^{\prime \prime}_{ijk} 
\hat{L}_{i}\hat{Q}_{j}\hat{d}^{c}_{k} \right) \,.
\label{superpotential1}
\end{eqnarray}
In the above equation, we have introduced two new terms that explicitly break the $R$-parity and the lepton number symmetry. There is a new parameter $\lambda^{\prime}$ that generates one more contribution to the mixing between the usual leptons with higgsinos. The usual techniques allow to determine from the superpotential $W$ the following mass matrix elements 
\begin{eqnarray}
&-&\left[ f^{l}_{ij} \left( H_{1}L_{i}l^{c}_{j}+ \tilde{H}_{1}\tilde{L}_{i}l^{c}_{j}\right) +
f^{\nu}_{ij}\left( \tilde{H}_{2}L_{i}\tilde{\nu}^{c}_{j}+H_{2}L_{i}\nu^{c}_{j}+ \tilde{H}_{2}\tilde{L}_{a}\nu^{c}_{b} \right)  \right. \nonumber \\ 
&+& \left. h_{i}^{\nu}\left( \tilde{H}_{1}\tilde{H}_{2}\tilde{\nu}^{c}_{i}+H_{1}\tilde{H}_{2}\nu^{c}_{i}+ \tilde{H}_{1}H_{2}\nu^{c}_{i}\right) + \kappa_{ijk}\tilde{\nu}^{c}_{i}\nu^{c}_{j}\nu^{c}_{k}+2 \lambda^{\prime}_{ijk}\tilde{L}_{i}L_{j}l^{c}_{k} \right] \,\ . \nonumber \\
\end{eqnarray}
The terms that describe the mixing between the usual leptons with the gauginos are the same as in the MSSM. In our notation, they are given by the following relations
\begin{eqnarray}
\imath \sqrt{2}g\left[ \bar{L}_{i} \left( \frac{\sigma^{a}}{2} \right) \overline{\tilde{W}^{a}} \tilde{L}_{i} -
\overline{\tilde{L}}_{i}\left( \frac{\sigma^{a}}{2} \right) \tilde{W}^{a} L_{i} \right]  
- \imath \sqrt{2} \left[ \bar{L}_{i} \left( - \frac{1}{2} \right) \tilde{L}_{i} 
\overline{\tilde{V}^{\prime}}- \overline{\tilde{L}}_{i}\left( - \frac{1}{2} \right)L_{i}\tilde{V}^{\prime} \right]. 
\end{eqnarray}
The mixing between the usual leptons with the higgsinos is given by the equation
\begin{eqnarray}
&-&f^{l}_{ij}\tilde{H}_{1}\tilde{L}_{i}l^{c}_{j} +
f^{\nu}_{ij}\left( \tilde{H}_{2}L_{i}\tilde{\nu}^{c}_{j}+ \tilde{H}_{2}\tilde{L}_{i}\nu^{c}_{j} \right) +
h_{i}^{\nu}\left( H_{1}\tilde{H}_{2}\nu^{c}_{i}+ \tilde{H}_{1}H_{2}\nu^{c}_{i}\right)  .
\end{eqnarray}
Beside the new mixing sectors given above, there are interactions between gauginos and higgsinos given by the same terms as in the MSSM. One can calculate the mass matrices of the charged leptons following the reference \cite{Escudero:2008jg}. The result in the basis $\Psi^{- \,\ T}=(-i \tilde{W}^{-},\tilde{H}^{-}_{1},l_{1},l_{2},l_{3})^{T}$ is given by the 
matrix
\begin{equation}
M_{C}= \frac{1}{\sqrt{2}}
\left(
\begin{array}{ccccc}
\sqrt{2}M_{2} & gv_{2} & 0 & 0 & 0\\
gv_{1} & \lambda_{i} v_{\nu^{c}_{i}} & -f^{l}_{i1} v_{\nu_{i}} & 
-f^{l}_{i1} v_{\nu_{i}} & -f^{l}_{i1} v_{\nu_{i}} \\
gv_{\nu_{1}} & -f^{\nu}_{1i}v_{\nu^{c}_{1}} & a_{11} &  a_{12} &  a_{13} \\
gv_{\nu_{2}} & -f^{\nu}_{2i}v_{\nu^{c}_{2}} &  a_{21} & a_{22} &  a_{23} \\
gv_{\nu_{3}} & -f^{\nu}_{3i}v_{\nu^{c}_{3}}&  a_{31} &  a_{32} & a_{33}
\end{array}
\right),
\label{charginosmasses}
\end{equation}
where
\begin{equation}
a_{ij}= f^{l}_{ij}v_{1}-\sum_{k=1}^{3}\lambda^{\prime}_{kij} v_{\nu_{k}}\sim f^{l}_{ij}v_{1}.
\end{equation}
Here, the winos $\tilde{W}^{\pm}$ (superpartners of the $W$-boson) defined as
\begin{equation}
\sqrt{2}\tilde{W}^{\pm}\equiv \tilde{V}^{1}\mp \tilde{V}^{2} \, ,
\label{Winos-def}
\end{equation} 
See also the equation (\ref{vevemunussm}). 
In the neutralino sector we use the basis 
\begin{equation}
\Psi^{0 \,\ T}=(-i \tilde{V}^{\prime},-i \tilde{V}^{3},\tilde{H}^{0}_{1},\tilde{H}^{0}_{2},\nu^{c}_{1},\nu^{c}_{2},\nu^{c}_{3},\nu_{1},\nu_{2},\nu_{3})^{T}.
\label{neutralino-basis}
\end{equation}
Then the mass matrices take the following form
\begin{equation}
M_{N}= 
\left(
\begin{array}{cc}
M_{7 \times 7} & m_{3 \times 7} \\
(m_{3 \times 7})^{T} & 0_{3 \times 3}
\end{array}
\right)_{10 \times 10},
\label{neutralinosmasses}
\end{equation}
where $M_{N}$ is the neutralino mass matrix presented in \cite{Escudero:2008jg}. However, one should emphasize that the neutrinos are Majorana particles. Therefore, the double beta decay without neutrinos can occur in this model. This process is permitted by the new first term from the equation (\ref{superpotential1}). The second term alone will not induce neither the fast proton decay, nor the neutron anti-neutron oscillation \cite{dress}.

It is important to note that the last term in the superpotential will modify the down quarks masses. Some algebra shows that they are given by the following relation
\begin{eqnarray}
m_{d}= \frac{1}{\sqrt{2}} \left(f^{d}v_{1}+\lambda^{\prime}_{ijk} v_{\nu_{i}} \right) \sim 
\frac{1}{\sqrt{2}}f^{d}v_{1}. 
\label{massa do lepton}
\end{eqnarray}
The new superpotential given by the equation (\ref{superpotential1}) has all properties required in \cite{LopezFogliani:2005yw,Escudero:2008jg,Munoz:2009an}. It has the advantage that it induces the double beta decay while maintaining the nucleon stability \cite{dress} as was discussed in the previous section.

\subsection{Flat direction of $\mu \nu$SSM Model}
\label{sec:cosmological}

One of the most remarkable aspects of the supersymmetric gauge theories is that they have a vacuum degeneracy at the classical level. It is a well established fact that the renormalizable scalar potential is a sum of squares of $F$-terms and $D$-terms which implies that it can vanish identically along certain \emph{flat directions} \footnote{The flat directions are noncompact lines 
and surfaces in the space of scalars fields along which the scalar potential vanishes. The present flat direction is an accidental feature of the classical potential and gets removed by quantum corrections.} in the space of fields
\footnote{In quantum field theories, the possible vacua are usually labelled by the vacuum expectation values of scalar fields, as Lorentz invariance forces the vacuum expectation values of any higher spin fields to vanish. These vacuum expectation values can take any value for which the potential function is a minimum. Consequently, when the potential function has continuous families of global minima, the space of vacua for the quantum field theory is a manifold (or orbifold), usually called the \emph{vacuum manifold}. This manifold is often called the moduli space of vacua, or just the moduli space.}
. The properties of the space of flat directions of a supersymmetric model are crucial for making realistic considerations in cosmology and whenever the behaviour of the theory at large field strengths is an issue.

In the MSSM the flat directions can give rise to a host of cosmologically
interesting dynamics (for a review, see \cite{Enqvist:2003gh}). These
include Affleck-Dine baryogenesis
\cite{Affleck:1984fy,Dine:1995uk,Dine:1995kz}, the cosmological
formation and fragmentation of the MSSM flat direction condensate and
subsequent $Q$-ball formation
\cite{Kusenko:1997zq,Enqvist:1997si,Jokinen:2002xw,Kasuya:1999wu},
reheating the Universe with $Q$-ball evaporation~\cite{Enqvist:2002rj},
generation of baryon isocurvature density perturbations
\cite{Enqvist:1998pf}, as well as curvaton scenarios where MSSM flat directions
reheat the Universe and generate adiabatic density perturbations
\cite{Enqvist:2002rf}. Adiabatic density perturbations
induced by fluctuating inflaton-MSSM flat
direction coupling  has also been discussed in \cite{Enqvist:2003uk}.

We can calculate all flat directions in the MSSM using the prescription given in  \cite{Dine:1995kz,Gherghetta:1995dv}. It turns out that a MSSM flat direction is some linear combination of the MSSM scalars and can be thought of as a trajectory in the moduli space described by a single
scalar degree of freedom.
\begin{table}[htb]
\renewcommand{\arraystretch}{1.10}
\begin{center}
\normalsize
 \vspace{0.5cm}
\begin{tabular}{|c|c|}
\hline
\hline
Flat direction & $(B-L)$ \\
\hline
\hline
$\hat{H}_{2}\hat{Q}\hat{u}^{c}$ & 0 \\
$\hat{H}_{1}\hat{Q}\hat{d}^{c}$ & 0 \\
$\hat{H}_{1}\hat{L}\hat{l}^{c}$ & 0 \\
$\hat{H}_{2}\hat{L}\hat{\nu}^{c}$ & 0 \\
$\hat{H}_{1}\hat{H}_{2}\hat{\nu}^{c}$ & 1 \\
$\hat{\nu}^{c}\hat{\nu}^{c}\hat{\nu}^{c}$ & 3 \\
$\hat{L}\hat{L}\hat{e}^{c}$ & -1 \\
$\hat{L}\hat{Q}\hat{d}^{c}$ & -1 \\
\hline
\hline
\end{tabular}
\caption{\small Flat direction of the model $\mu \nu$SSM.}
\label{tab:flatdir} 
\end{center}
\end{table}
Using the same technique from \cite{Dine:1995kz,Gherghetta:1995dv} we can calculate the flat directions in $\mu \nu$SSM model. The Table (\ref{tab:flatdir}) presents the computed flat directions when the only terms taken into account are the renormalizable terms. In this model 
$\hat{H}_{2}\hat{L}_{i}\hat{\nu}^{c}_{j}$ gives a particular flat direction. It follows that the field
\begin{equation}
\phi_1 = \frac{H_{2}+ \tilde{L}+ \tilde{\nu}^{c}}{\sqrt{3}},
\label{flat1}
\end{equation}
generates a  similar inflanton scenario as the one discussed in \cite{Allahverdi:2007wt,Allahverdi:2006cx,Allahverdi:2009kr}. Another flat direction is given by the interaction term $\hat{\nu}^{c}_{i}\,\hat{H}_{1} \hat{H}_{2}$ from which is generated the following field
\begin{equation}
\phi_2 = \frac{H_{1}+H_{2}+ \tilde{\nu}^{c}}{\sqrt{3}}.
\label{flat2}
\end{equation} 
It will be interesting to compare the cosmological consequences of the fields $\phi_1$ and $\phi_2$ given above.

\subsection{Leptogenesis in $\mu \nu$SSM Model}
\label{sec:leptom1}

The mechanisms to create a baryon asymmetry from an initially symmetric state must in general satisfy the three basic conditions for baryogenesis as pointed out by Sakharov~\cite{sakharov}: 
\begin{enumerate}
 \item Violate baryon number, $B$, conservation.
 \item Violate $C$ and $CP$ conservation.
 \item To be out of thermal equilibrium.
\end{enumerate}
It is found that the $CP$ violation observed in the quark sector \cite{KM_CPviolation}, e.g. in 
$K^0$-$\bar{K}^0$ or $B^0$-$\bar{B}^0$ mesons system, is far too small 
to give rise to the observed baryon asymmetry \cite{CP_in_SM_tooSmall}. Therefore, these conditions should be extended to include the lepton number $L$ violation processes.

The classic leptogenesis scenario of Fukugita and Yanagida \cite{fukugita_yanagida_86} described in \cite{LopezFogliani:2005yw,Escudero:2008jg} can occur in the $\mu\nu$SSM model \cite{Chung:2010cd}. The Yukawa coupling can then induce heavy right handed neutrino $N$ decays via 
the following two channels:
\begin{equation}
\label{eq1:lepto_std_decay_channels}
 N_k \rightarrow
 \left\{ 
  \begin{array}{r}
  \;\;\; l_{j} + \overline{\phi} \;,\\
  \;\;\; \overline{l}_{j} + \phi \;, 
  \end{array} 
 \right.
,
\end{equation}
that violate the lepton number by one unit. In the new $\mu\nu$SSM model given by the superpotential (\ref{superpotential1}), it is one of the heavies neutralinos that is responsible for the right handed neutrion decay according to the equation (\ref{neutralinosmasses}).
All Sakharov's conditions for leptogenesis are satisfied if these decays violate $CP$ and go out of equilibrium at some stage during the evolution of the early universe. The requirement for $CP$ violation means that the coupling matrix $Y$ must be complex and the mass of $N_k$ must 
be greater than the combined mass of $l_{j}$ and $\phi$, so that the interferences between the tree-level processes and the one-loop 
corrections with on-shell intermediate states will be non-zero 
\cite{Law:2009vh,Law:2010zz}. 
Since $\phi$ is the scalar field of the SM, the usual Higgs can suffer the following decays
\begin{eqnarray}
H^{0}_{1} & \rightarrow & l_{a}l^{c}_{b},
\\
H^{0}_{2} & \rightarrow & \nu_{a}\nu^{c}_{b},
\\
H^{-}_{1} & \rightarrow & \nu_{a}l^{c}_{b} ,
\\ 
H^{+}_{2} & \rightarrow & l_{a}\nu^{c}_{b}.
\end{eqnarray} 
Note that none of these  decays violate the lepton number conservation. Nevertheless, in this model the fields $\tilde{\nu}$ have both chiralities. Therefore, they will induce the followings decays
\begin{eqnarray}
\tilde{\nu}^{c}_{c} & \rightarrow & \nu^{c}_{a}\nu^{c}_{b},
\\
\tilde{\nu}_{a} & \rightarrow & l_{b}l^{c}_{c}.
\end{eqnarray}
Thus, both violate the lepton number conservation.
On the other hand, we note that there are scattering processes that can alter the abundance of the neutrino flavour $N_K$ in the $s$-channel 
$N \ell \leftrightarrow q_L \bar{t}_R$ and $t$-channel
$N t_R \leftrightarrow q_L \bar{\ell}, N q_L \leftrightarrow t_R \ell$ besides the tree-level interaction ($N\leftrightarrow \ell \bar{\phi}$). In addition to these, there are also $\Delta L = \pm 2$ scattering processes mediated by $N_k$ which can be important for the evolution of $(B-L)$. Also, if we consider the couplings $Y_{\nu}^{ij}$ and $\lambda^{i}$ to be complex, we can generate 
the leptogenesis in this model as shown in \cite{Law:2009vh} by inducing decays as 
$\tilde{\chi}^{0}l \rightarrow d\bar{u}$.

It is interesting to note that the superpotential from the equation (\ref{superpotential1}) induces the following processes \cite{dress,Baer:2006rs,barbier,moreau}
\begin{enumerate}
\item New contributions to the neutrals $K\bar{K}$ and 
$B\bar{B}$ Systems.
\item New contributions to the muon decay.
\item Leptonic Decays of Heavy Quarks Hadrons such as 
$D^{+} \rightarrow \overline{K^{0}}l^{+}_{i}\nu_{i}$.
\item Rare Leptonic Decays of Mesons like $K^{+} \rightarrow \pi^{+}\nu \bar{\nu}$.
\item Hadronic $B$ Meson Decay Asymmetries.
\end{enumerate}
Also, ir gives the following direct decays of the lightest neutralinos
\begin{eqnarray}
\tilde{\chi}^{0}_{1} &\rightarrow& l^{+}_{i}\bar{u}_{j}d_{k}, \,\
\tilde{\chi}^{0}_{1} \rightarrow l^{-}_{i}u_{j}\bar{d}_{k},
\nonumber \\
\tilde{\chi}^{0}_{1} &\rightarrow& \bar{\nu}_{i}\bar{d}_{j}d_{k},
\,\ \tilde{\chi}^{0}_{1} \rightarrow \nu_{i}d_{j}\bar{d}_{k},
\label{lepto1}
\end{eqnarray}
and for the lightest charginos
\begin{eqnarray}
\tilde{\chi}^{+}_{1} &\rightarrow& l^{+}_{i}\bar{d}_{j}d_{k}, \,\
\tilde{\chi}^{+}_{1} \rightarrow l^{+}_{i}\bar{u}_{j}u_{k},
\nonumber \\
\tilde{\chi}^{+}_{1} &\rightarrow& \bar{\nu}_{i}\bar{d}_{j}u_{k},
\,\ \tilde{\chi}^{+}_{1} \rightarrow \nu_{i}u_{j}\bar{d}_{k}.
\label{lepto2}
\end{eqnarray}
These decays are similar to the ones from the MSSM when $R$-Parity violating scenarios are taken into account. Therefore, we expect that the missing energy plus jets be the main experimental signal in the 
"new" $\mu \nu$SSM as is in the MSSM. These decays violate only the lepton number conservation but they conserve the baryon number.

As we have seen above, all necessary conditions to generate a viable leptogenesis mechanism from the $\mu \nu$SSM model are present in the "new" $\mu \nu$SSM model \cite{Kajiyama:2009ae} as well as the CP violation processes. Also, this model could contain an invisible axion. These properties deserve a deeper study. Another interesting phenomenological avenue is to analyse the total cross section of the Dark Matter-Nucleon (DM-N) elastic scattering process. 

\section{Explanation of the data from ATLAS, CMS and LHCb in $\mu \nu$SSM Model.}
\label{sec:data}

One possible explanation to the excess of electrons is given if the following processes are considered \cite{Biswas:2014gga,Allanach:2014lca}
\begin{eqnarray}
pp & \rightarrow & \tilde{e} \rightarrow e^{-} \tilde{\chi}_{1}^{0} \rightarrow e^+e^-jj, \nonumber \\
pp & \rightarrow & \tilde{\nu_{e}} \rightarrow e^{-} \tilde{\chi}_{1}^{+} \rightarrow e^+e^-jj.
\label{cmsexplanation}
\end{eqnarray}
Neglecting finite width effects, the color and spin-averaged parton total cross section of a single slepton production is~\cite{Dimopoulos:1988jw,Allanach:2009iv}
\begin{eqnarray}
\hat{\sigma} &=& \frac{\pi}{12\hat{s}}|\lambda^{\prime}_{111}|^2\delta \left( 1-\frac{m^{2}_{\tilde{l}}}{\hat{s}} \right),
\end{eqnarray}
where $\hat{s}$ is the partonic center of mass energy, and $m_{\tilde{l}}$ is the mass of the resonant slepton. Including the effects of the parton distribution functions, we find the total cross section
\begin{eqnarray}
\sigma (pp \rightarrow \tilde{l}) \propto |\lambda^{\prime}_{111}|^{2} / m_{\tilde{l}}^{3}, 
\end{eqnarray}
to a good approximation in the parameter region of interest.

As was discussed in \cite{Allanach:2014lca}, these processes represent one of the possibilities to  explain the data of CMS  \cite{Khachatryan:2014dka,CMS1} if the selectron mass is fixed to $2.1 {\mbox TeV}$ and the lightest neutralino mass is taken to be in the range from 400 GeV up to 1 TeV. The $R_{K}$ measurement can be consistent with the new physics arising from the electron or muon sector of the SM and it was shown in \cite{Biswas:2014gga} that if we consider the muon sector in the MSSM with $R$-parity violation scenarios, the $R_{K}$ can also account for both data arising from CMS and LHCb. In the "new" $\mu\nu$SSM model we have both terms present. 
With respect with the di-boson data, there is a similar explanation. Indeed, in the case of $V=W,Z$ there is the single 
production of smuons \cite{Allanach:2015blv}, while in the case of di-photons the stau is produced \cite{Allanach:2015ixl}. Due this fact, we expect that our model fit the new data coming from ATLAS \cite{atlas13}, CMS \cite{CMS1} and from LHCb \cite{Aaij:2014ora,Aaij:2013pta}. To confirm that this is the true mechanism employed, the double beta decay must be detected in experiments like CUORE \cite{CUORE}, GERDA \cite{GERDA} and MAJORANA \cite{MAJORANA} and no proton decay must occur in the neutron anti-neutron oscillation.

\section{Conclusions}
\label{sec:conclusion}

In this article we have reviewed some of the basic properties of the MSSM, NMSSM and $\mu \nu$SSM essential to the cosmological applications. Also, in order to incorporate the recent data from the CMS and LHCb into this class of models, we have proposed a "new" $\mu \nu$SSM model characterized by the superpotential given in the equation (\ref{superpotential1}). The terms added to the $W_{superpot}$ of the $\mu\nu$SSM in order to obtain the modified model, explicitly break the $R$-parity and the lepton number conservation. This makes the model attractive for cosmological applications as it presents flat directions that represent a possibility to generate inflation and a viable leptogenesis mechanism that is necessary to generate the matter anti-matter asymmetry. These properties make the model interesting for further investigations on which we hope to report in the near future.

\begin{center}
{\bf Acknowledgments} 
\end{center}
M. C. R would like to thanks to Laborat\'{o}rio de F\'{\i}sica Experimental at 
Centro Brasileiro de Pesquisas F\'\i sicas (LAFEX-CBPF) for their nice hospitality and  
special thanks to Professores J.A. Helay\"{e}l-Neto and A. J. Accioly. Both authors acknowledge R. Rosenfeld for hospitality at ICTP-SAIFR where part of this work was accomplished. We acknowledge P. S. Bhupal Dev for information 
on the latest results from ATLAS and CMS in di-bosons, and D. E. Lopez-Fogliani for useful correspondence on the gravitino in the $\mu \nu$SSM.


\end{document}